\RequirePackage[l2tabu]{nag}
\documentclass{article}

\usepackage{fullpage}
\usepackage{float}
\usepackage{graphicx}   
\usepackage{wrapfig}
\usepackage{caption}
\usepackage{subcaption}
\usepackage{courier}
\usepackage{enumitem}
\usepackage{soul,color}
\usepackage{balance}
\usepackage{multirow}

\usepackage[utf8]{inputenc}
\usepackage[T1]{fontenc}

\usepackage{lmodern}
\usepackage[tracking=true,kerning=true,final]{microtype}
\usepackage{bbm}

\usepackage{bm}

\usepackage{url}
\usepackage[unicode=true,pdfusetitle, bookmarks=true,bookmarksnumbered=false,bookmarksopen=false,
breaklinks=false,hidelinks,colorlinks=true]{hyperref}
\hypersetup{colorlinks=true, linkcolor=blue, citecolor=blue, urlcolor=blue}

\usepackage{soul,color}
\usepackage[utf8]{inputenc}
\usepackage[english]{babel}

\usepackage{hhline}

\usepackage[utf8]{inputenc}
\usepackage[T1]{fontenc}

\usepackage{color}

\usepackage{lmodern}
\usepackage[tracking=true,kerning=true,final]{microtype}

\title{Bubbles in Turbulent Flows: Data-driven, kinematic models with history terms}
\author{Zhong Yi Wan$^1$, Petr Karnakov$^2$, Petros Koumoutsakos$^2$, Themistoklis P. Sapsis$^1$
\thanks{Corresponding author: \href{mailto:sapsis@mit.edu}{sapsis@mit.edu},
Tel: (617) 324-7508, Fax: (617) 253-8689%
}\\
$^1 $Department of Mechanical Engineering, Massachusetts Institute of Technology, \\
$^2 $ Computational Science and Engineering Laboratory, ETH Z\"urich, Switzerland
}
\date{\today}

\usepackage{amsmath}
\usepackage{amssymb}

\usepackage[normalem]{ulem}

\begin{document}

\maketitle 
\begin{abstract}
We present data driven  kinematic models for the motion of bubbles in high-Re turbulent fluid flows based on recurrent neural networks with long-short term memory enhancements. The models extend empirical relations, such as Maxey-Riley (MR) and its variants, whose applicability is limited when either the bubble size is large or the flow is very complex. The  recurrent neural networks are trained on the trajectories of bubbles obtained by Direct Numerical Simulations (DNS) of the Navier Stokes equations for  a two-component incompressible flow model. Long short term memory components exploit the time history of the flow field that the bubbles have encountered along their trajectories and the networks are further  augmented by imposing rotational invariance to their structure. We first train and validate the formulated model using DNS data for a turbulent Taylor-Green vortex. Then we examine the model predictive capabilities and its generalization to Reynolds numbers that are different from those of the training data on benchmark problems, including a steady (Hill's spherical vortex) and an unsteady (Gaussian vortex ring) flow field. We find that the predictions of the developed model are significantly improved compared with those obtained by the MR equation. Our results indicate that data-driven models with history terms are well suited in capturing the trajectories of bubbles in turbulent flows.
  
\end{abstract}

\section{Introduction}
Inertial particles, such as bubbles or aerosols, have been the focal point of a large number of studies, from particle ladden turbulent flows \cite{elghobashi2019a} to cloud dynamics \cite{Shaw2003}, over the last few decades. These  studies reveal that even in the case of neutral buoyancy, inertial particles may not follow the flow, exhibiting strongly dispersive or clustering behaviors \cite{maxey83, michaelides97, rubin95}. In this context several dynamical models have been derived for the description of inertial particles motion.  

The theoretical analysis of inertial particle motion was pioneered by G. Stokes \cite{stokes1851}, who addressed the motion of an isolated  particle sedimenting in a fluid when the inertia effects are  negligible and the flow field is dominated by viscous diffusion. Since then numerous efforts have been made to derive improved models (see e.g. \cite{basset1888, oseen1910, proudman, sano}). Maxey and Riley \cite{maxey83} carried out a complete analysis of the motion of a sphere in unsteady Stokes flow for nonuniform flow fields and derived an  equation governing the relative velocity of the particle for any nonuniform transient background flow. This model and several of its improvements have been used widely for the description of bubbles, aerosols and neutrally buoyant particles in ambient fluid flows.

The Maxey-Riley (MR) equation is characterized by a singular structure and is challenging to solve numerically. In \cite{Haller2008} a geometrical singular perturbation reduction was performed for small particles size that led to analytical kinematic models for inertial particles. These models were employed to investigate properties such as dispersion \cite{Sapsis2008a, Haller2010} and clustering \cite{Sapsis2010}. Despite the success of the MR equation (and its variants) on qualitatively capturing several aspects of inertial particles motion, there are still significant discrepancies between the predicted trajectories of individual particles and those measured experimentally, even for relatively low Re numbers \cite{sapsis2011IP}.  

In \cite{wan_sapsis_2018} it was shown that bubble trajectories generated from the MR equation for a simple cell flow, are sufficient to facilitate the learning of data-driven kinematic models. These models not only reproduce  the trajectories for the flow in which they were trained, but they also predict accurately bubble trajectories and their non-Gaussian statistics in different flows, such as the flow behind a cylinder or a turbulent flow in a periodic box. This is achieved by learning a velocity field for the particles, in terms of the flow properties that the particle has encountered along its past trajectory. In this way the trained model does not depend on the specific form of the flow that is used for training. Despite its success, this study was limited by the fact that the underlying `ground truth' is considered to be the MR equation and the trajectories generated from it, which, due to a multitude of simplifying assumptions made (e.g. size and shape of the bubble), is a good assumption for a limited number of problems. In turn, direct numerical simulations (DNS), can provide more realistic descriptions of the underlying physical phenomena, as the whole range of spatial and temporal scales in the Navier-Stokes (NS) equations are resolved. DNS of  turbulent flows with bubbles, however, entail a  number of technical challenges, including the accurate description of the liquid-gas interface, the treatment of the surface tension force \cite{elghobashi2019a} and the high computational cost. 


The scope of this work is to obtain a data-driven kinematic model for 3-dimensional (3D) finite-size bubbles, trained using a limited number of \textit{DNS} bubble trajectories. Compared to prior work, the present  model is subject to fewer restricting assumptions as it learns from high-fidelity NS solutions and therefore it is  capable of providing accurate description of bubble behaviors under a much wider range of settings. In particular, we aim to learn kinematic models for bubbles that generalise well to other fluid flows, by informing  the  model with the local flow characteristics that the bubbles have encountered along their trajectories. At the same time, we address the data scarcity problem that arises as a natural consequence of relying on data generated by expensive DNS. Specifically, we exploit physical symmetries of the kinematic models, such as rotation invariance, in order to augment the limited training data set. We examine the  generalization  of the trained model to different Re numbers and  assess its performance in three different 3D benchmark problems. The first one is  the turbulent Taylor-Green vortex which also provides the training data. Subsequently, we use the same trained model for bubbles in a flow field described by the laminar Hill's spherical vortex, as well as, bubbles in a flow field of an unsteady vortex ring generated by DNS. Detailed comparisons with MR are also presented.

The remainder of this work is  organized as follows. In section \ref{sec:meth} we present the DNS set-up that is used to obtain bubble trajectories required for model training and validation. Section \ref{sec:model} provides details of the data-driven model including the selection of the appropriate input variables for the model, a novel data augmentation mechanism to deal with sparse data, and training results. Using the learned model we also examine its generalizability to different flow conditions, including different Reynolds (Re) numbers. Finally, we showcase the model capabilities to: (a, section \ref{sec:hills}) predict bubbles motion in an analytically described, steady, fluid flow, the Hill's spherical vortex, and (b, section \ref{sec:ring}) provide high-fidelity multi-step predictions for bubbles in an unsteady vortex ring.
    
\section{DNS of a turbulent multiphase Taylor-Green vortex}
\label{sec:meth}

The training data for our model is generated from DNS of turbulent flows with bubbles, which provide complete information about the bubbles trajectories and the underlying flow. We focus  on obtaining  kinematic models for monodisperse systems, i.e. mixtures with bubbles of equal size. The important case of polydisperse systems will be considered in the future. For DNS data generation, we employ a two-component incompressible flow model consisting of the 3D Navier-Stokes equations for the mixture velocity~$\textbf{u}$ and pressure~$p$
\begin{align}
        \nabla \cdot \textbf{u} &= 0, \\
        \label{eq:NS}
        \rho\Big(\frac{\partial \textbf{u}}{\partial t} + (\textbf{u} \cdot
\nabla)\, \textbf{u}\Big) &= -\nabla p + \nabla \cdot \mu (\nabla \textbf{u}
+ \nabla \textbf{u}^T ) + \textbf{f}_\sigma
\end{align}
and the advection equation for the volume fraction~$\alpha$
\begin{equation}
    \frac{\partial \alpha}{\partial t} + (\textbf{u} \cdot \nabla)\, \alpha = 0,
    \label{eq:volfrac}
\end{equation}
where the mixture density~$\rho=(1-\alpha)\rho_1 + \alpha\rho_2$ and dynamic viscosity~$\mu=(1-\alpha)\mu_1 + \alpha\mu_2$ are computed from the volume fraction and constant material parameters~$\rho_1$, $\rho_2$, $\mu_1$ and $\mu_2$. The surface tension force is defined as $\textbf{f}_\sigma = \sigma \kappa \nabla \alpha$ where $\sigma$ and $\kappa$ denote the surface tension coefficient and interface curvature respectively. The equations are discretized on a uniform Cartesian mesh using the finite volume method and the SIMPLE algorithm for pressure coupling \cite{patankar1983,ferziger2012}. The advection equation is solved using the volume-of-fluid method with piece-wise linear interface reconstruction~\cite{aulisa2007}. For a detailed description and evaluation of the numerical algorithm, including the treatment of surface tension, readers are referred to~\cite{partstr}.

\begin{figure}[t]
        \centering
        \includegraphics[width=0.32\textwidth,trim={50pt 0 50pt 0},clip]{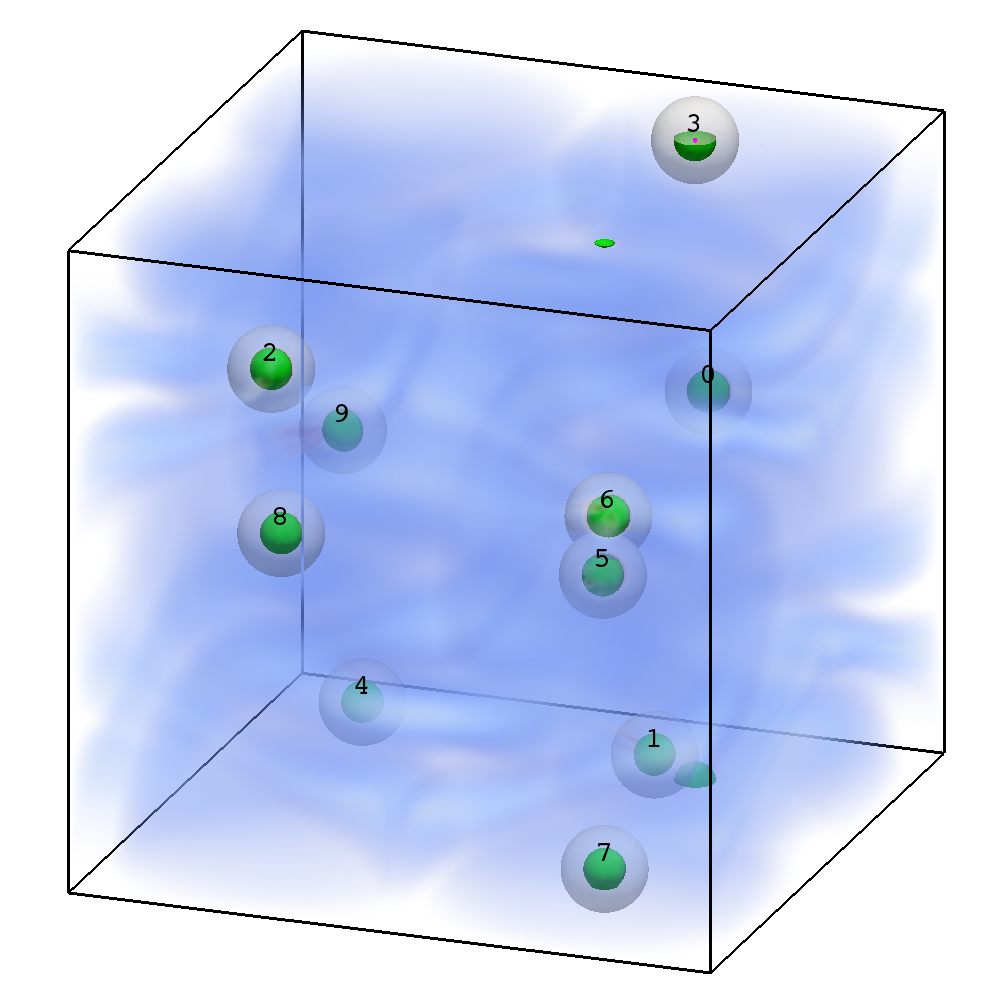}%
        \hfill
        \includegraphics[width=0.32\textwidth,trim={50pt 0 50pt 0},clip]{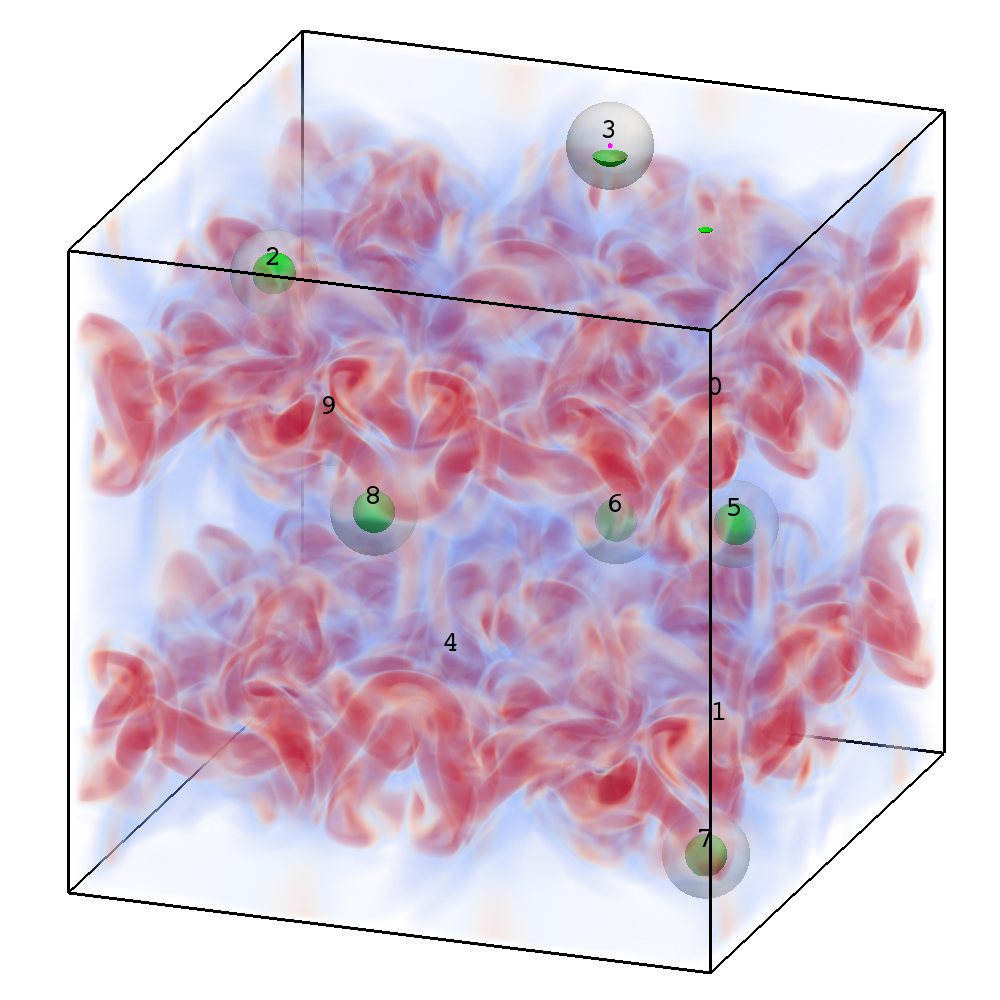}%
        \hfill
        \includegraphics[width=0.32\textwidth,trim={50pt 0 50pt 0},clip]{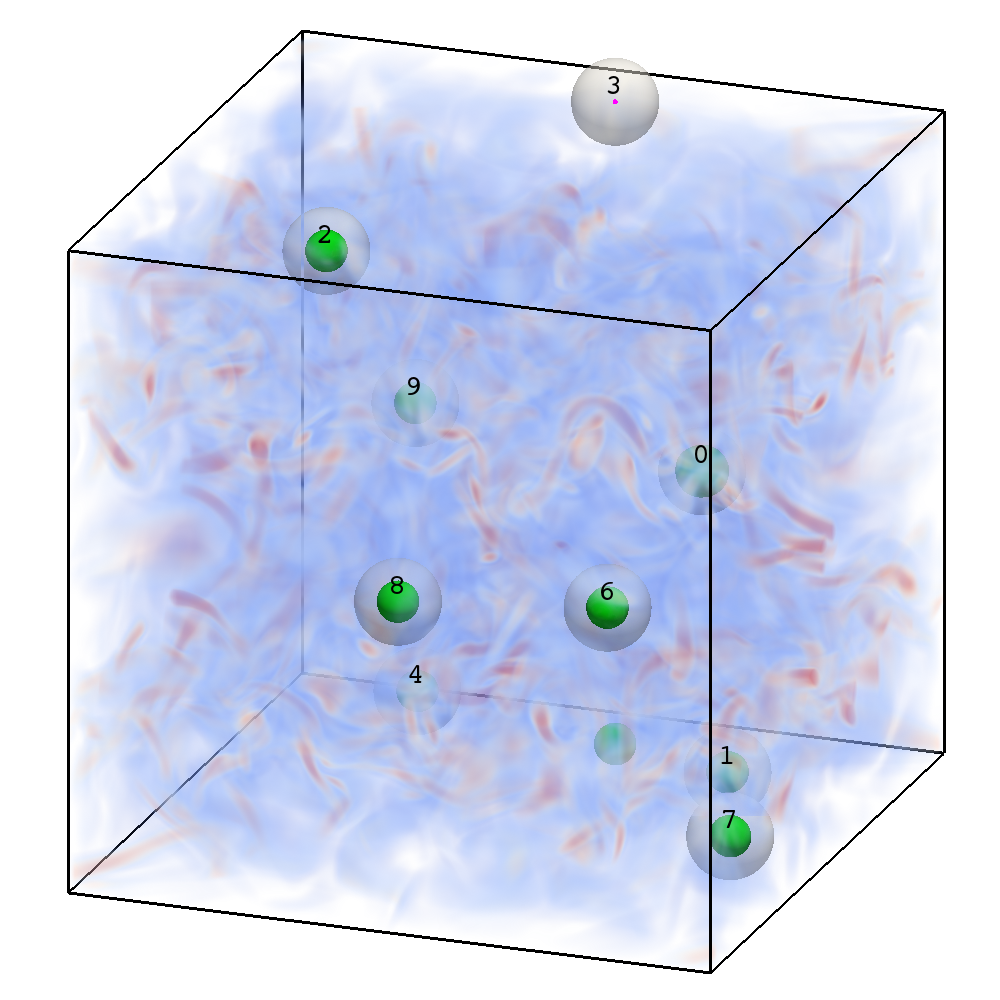}

        \caption{
          Snapshots of the DNS flow field for the multiphase Taylor-Green vortex with $\text{Re}=800$
          at time instants~$t=2.5,\;10,\;17.5$.
          The vorticity field is shown with color
          (increasing values from blue to red)
          together with the surfaces of bubbles (green) and their indices.
          Properties of the bubbles
          (such as position, pressure, velocity) 
          are obtained by averaging over spherical surfaces surrounding the bubbles (gray).}
        \label{fig:flow}
 \end{figure}

\begin{figure}[t]
        \centering
        \includegraphics{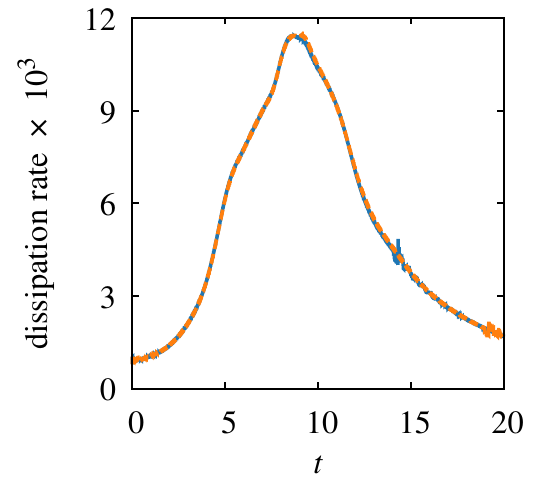}\hfill
        \includegraphics{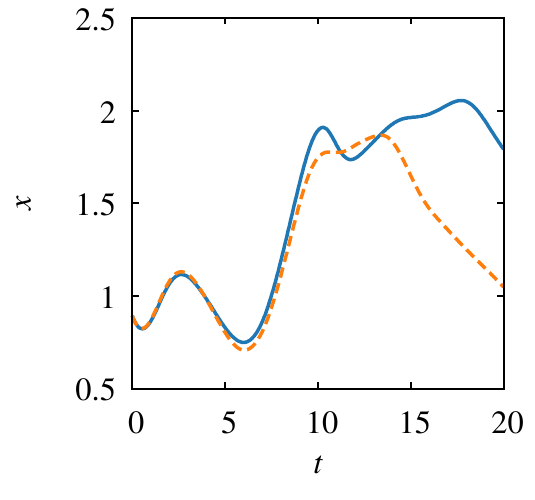}\hfill
        \includegraphics{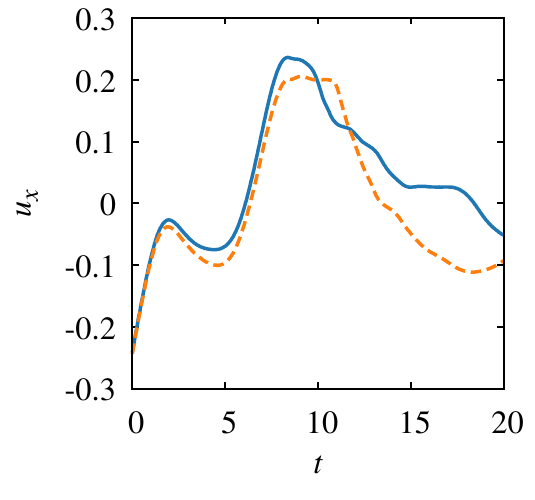}
        \caption{Energy dissipation rate 
        $\tfrac{1}{8\pi^3}\int\rho|\textbf{u}|dV$
        and 
        $x$-components of position and flow velocity 
        of bubble~4 from Figure~\ref{fig:flow}
        depending on the mesh size:
        $256^3$ (solid blue) and $384^3$ (dashed orange).
        }
        \label{fig:conv}
 \end{figure}

We generate data for the training of our model by simulating the  evolution of the Taylor-Green vortex ~\cite{rees2011} extended by a gaseous phase. The problem is solved in a periodic domain $\Omega = [0,2\pi]^3$ with the initial velocity field $(u_x, u_y, u_z) = (\sin x\cos y\cos z, -\cos x\sin y\cos z, 0)$ and the volume fraction field representing multiple bubbles of dimensionless radius $R=0.196$. The liquid has dimensionless density $\rho_1=1$ and viscosity $\mu_1=1/\text{Re}$ based on the Reynolds number, $\text{Re}$. For the gas properties, we assume that $\mu_2=0.01\,\mu_1$ and $\rho_2=0.01\,\rho_1$. The surface tension coefficient $\sigma = 2R/\text{We}$ is determined from the Weber number, $\text{We}$, which characterizes the ratio between inertia and surface tension. The trajectories of bubbles are obtained by computing the center of mass of each bubble from the volume fraction field. The flow properties along each of the bubbles trajectories are obtained by averaging the flow field over a spherical surface of radius $2R$ surrounding each bubble. The problem is solved on a mesh of~$256^3$ cells. Snapshots of the simulated flow field at $\text{Re}=800$ are shown in Figure~\ref{fig:flow}. Results in~Figure~\ref{fig:conv} show how the solution changes with mesh refinement: the integral quantities such as the energy dissipation rate are independent of the mesh size while the trajectories deviate at later times due to their chaotic nature.

The  data set includes three cases with $\text{Re}=400,\;800$ and $1600$ and in all cases $\text{We}=3.92$. For each Re, 30 groups of~10 bubbles are simulated. The bubbles are initialized with zero velocities and random positions drawn from a uniform distribution in the problem domain. Within each group, the bubbles may occasionally coalesce with one another to form a bigger bubble but maintain a constant size otherwise. For the purpose of this work, we are only considering bubbles with constant size and trajectories with coalescence are therefore omitted. We obtain  $255\sim290$ bubble trajectories  for each Re number. 

 \begin{figure}[t]
        \centering
        \includegraphics{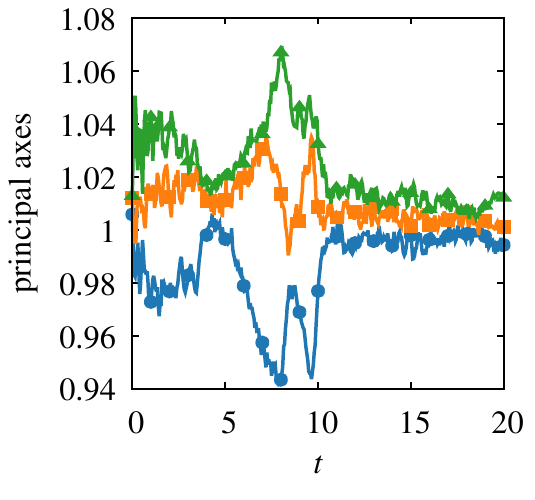}\hfill
        \includegraphics{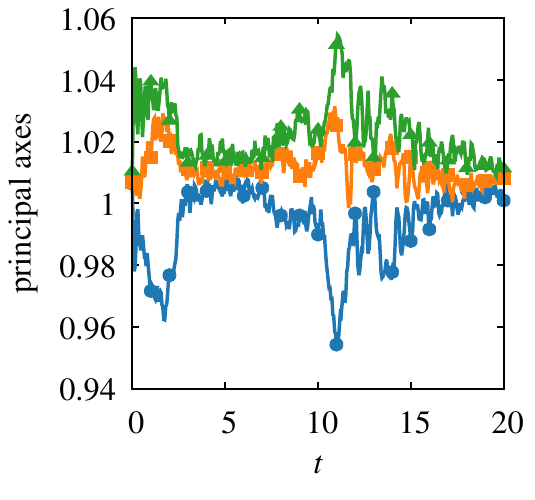}\hfill
        \includegraphics{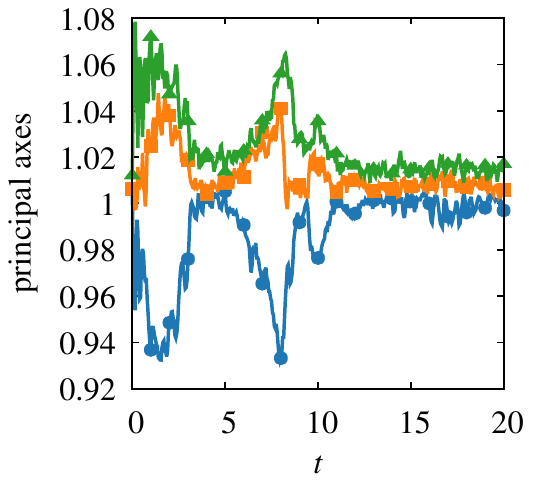}\hfill
        \caption{Principal axes of the gyration tensor of bubbles 
        2, 4 and 6 from Figure~\ref{fig:flow}
        normalized by the initial radius.  }
        \label{fig:deform}
 \end{figure}

\section{Data-driven modeling of bubbles in the Taylor-Green vortex}
\label{sec:model}

Following \cite{wan_sapsis_2018}, we aim to obtain a kinematic model relating the instantaneous velocity of the bubble to the history of the local flow conditions experienced by the bubble. As discussed, our study focuses on modeling the motion of bubbles whose radius stays constant ($R = 0.196$ unless otherwise noted). In discrete time, the model takes the form
    \begin{equation}
        \bar{\textbf{v}}_t = G\big(\boldsymbol{\Theta}; \hat{\boldsymbol{\xi}}_t, \hat{\boldsymbol{\xi}}_{t-1}, \hat{\boldsymbol{\xi}}_{t-2}, ...\big), 
        \label{eq:modeldata}
    \end{equation} 
where subscripts are used to represent the index of the time steps, $\bar{\textbf{v}}_t$ denotes the gas velocity averaged over the entire spherical volume of the bubble, and $\hat{\boldsymbol{\xi}}_t$ denotes the local
flow state (e.g. flow velocity $\textbf{u}$, gradient $\nabla \textbf{u}$, material derivative $D\textbf{u}/Dt$) experienced by the bubble and is an average value computed over the spherical bubble surface.

\begin{figure}[t]
        \centering
        \includegraphics[width=.35\linewidth]{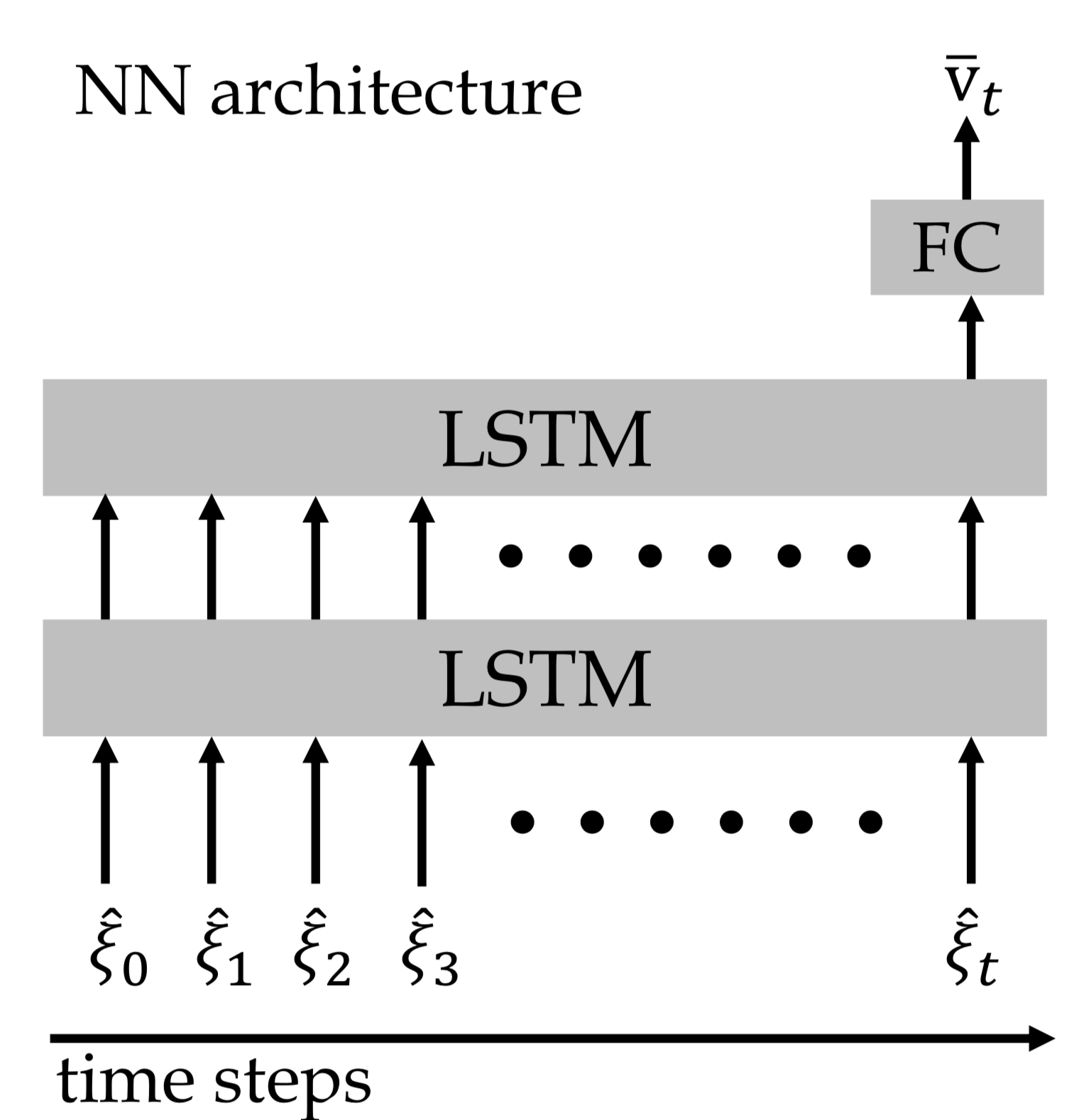}
        \caption{RNN/LSTM-based architecture for modeling bubble kinematics. The LSTM layers have 100 hidden units and fully-connected (FC) layer has 50. L2 kernel and recurrent regularizations are used ($\lambda=10^{-4}$).}
        \label{fig:arch}
    \end{figure}

The vector field, $G(\cdot)$, is represented with a recurrent neural network (RNN), whose structure is illustrated in Figure \ref{fig:arch}. RNNs have been shown to represent dynamical systems very effectively even for chaotic regimes \cite{Vlachas2018a}. The architecture consists of two stacked long short-term memory (LSTM) layers of 100 hidden units each and a fully-connected (FC) layer with 50 hidden units before the final 3-dimensional output layer representing bubble velocity. Denoting a set of $N$ training trajectories by $\mathcal{D} = \{\mathcal{T}^{(n)}\}_{n=1}^N$ where $\mathcal{T}^{(n)}=\{\bar{\textbf{v}}_1^{(n)}, \hat{\boldsymbol{\xi}}_1^{(n)},...,\bar{\textbf{v}}_T^{(n)}, \hat{\boldsymbol{\xi}}_T^{(n)}\}$ is a single training trajectory, the weights $\boldsymbol{\Theta}$ for RNN
model $G$ are obtained by minimizing the normalized mean squared error (MSE) loss function:
    \begin{equation}
        L(\boldsymbol{\Theta}; \mathcal{D}) = \frac{1}{N}\frac{1}{T}\sum_{n=1}^N\sum_{t=1}^T \frac{\left|\left|G\left(\boldsymbol{\Theta}; \hat{\boldsymbol{\xi}}^{(n)}_t, \hat{\boldsymbol{\xi}}^{(n)}_{t-1}, ...\right) - \bar{\textbf{v}}_t^{(n)}\right|\right|^2}{\text{var}(v)},
    \end{equation} 
where $\text{var}(v)$ is the variance of the bubble velocity.

\subsection{Maxey-Riley (MR) slow manifold approximation}
 
As a benchmark, we employ the MR equation:
    \begin{equation}
        \dot{\textbf{v}} = \frac{1}{\epsilon}(\textbf{u}-\textbf{v}) + \frac{3\gamma}{2}\frac{D\textbf{u}}{Dt},
        \label{eq:MRfull}
    \end{equation} 
where $\epsilon$ is a dimensionless parameter representing bubble inertia and $\gamma=2\rho_1/(\rho_1+2\rho_2)$ is a density ratio. The MR equation is known to admit a slow-manifold approximation which takes the asymptotic form  \cite{Haller2008}
    \begin{equation}
        \tilde{\textbf{v}} = \textbf{u} + \epsilon\left[\frac{3\gamma}{2}-1\right]\frac{D\textbf{u}}{Dt}+O(\epsilon^2),
        \label{eq:MRsm}
    \end{equation}  
which is a good approximation to the original equation when $\epsilon$ is small and some additional stability constraint (based on the minimum eigenvalue of $\nabla\tilde{\textbf{v}}$) is satisfied \cite{Sapsis2008a}. For the DNS parameters in \ref{sec:meth}, $\epsilon$ takes the value 3.48 at $\text{Re}=800$. This suggests that the bubble has considerable inertia, which significantly compromises the accuracy of the low-order truncation of the slow manifold (\ref{eq:MRsm}) and makes it difficult to assess its stability. Nevertheless, we  employ this model as a benchmark to understand the capabilities of our proposed machine-learning schemes. The resulting first order truncation to (\ref{eq:MRsm}) leads to standardized MSE of 0.77 and 0.83 respectively.

\subsection{Selection of input variables for the data driven, kinematic model}
 
In data driven models a  key issue is the selection of the flow features or variables that should be included as input, i.e. how to select the components of the input vector $\boldsymbol{\xi}$. Unlike the cases where the MR equation is valid \cite{wan_sapsis_2018}, in the present setup the form or even the presence of a slow manifold is not supported by any theoretical argument. This  opens up the number of candidate variables that could lead to good models, e.g. flow velocity, gradient of pressure, etc. 

We consider several combinations of input variables (see Appendix A) and observe that the 6-dimensional input vector $\hat{\boldsymbol{\xi}}$ consisting of the flow velocity $\textbf{u}$ and the flow material derivative $D\textbf{u}/Dt$ produces the best model performance after 100 epochs of training. We find that the flow material derivative is  seemingly playing a more important role. Including additional variables in the state vector slows down the training process and results in  a larger generalization gap between and training and validation losses. We emphasize that these particular variables are consistent with the ones contained in the standard form of the MR equation.

\subsection{Data augmentation via random orthogonal transformation}

The generation of DNS data has a high computational cost and it inevitably leads to a data scarcity problem. We propose a remedy to this situation by introducing a data augmentation scheme that exploits symmetries in the system to enrich our training data. In particular, due to the assumption that bubbles remain spherical at all times, the resulting kinematic model should be equivariant with respect to the orientation of the coordinate axes, i.e.
    \begin{equation}
        \mathcal{R}\bar{\textbf{v}}_t = G\big(\boldsymbol{\Theta}; \mathcal{R}\hat{\boldsymbol{\xi}}_t,\mathcal{R}\hat{\boldsymbol{\xi}}_{t-1}, \mathcal{R}\hat{\boldsymbol{\xi}}_{t-2},...\big), 
    \end{equation}
where $\mathcal{R}$ denotes an arbitrary orthogonal transformation. To facilitate training models with such a property,  we randomly sample orthogonal matrices which are used to transform the original data. In this way, we obtain an augmented training data set that is used to impose the directionless property on the model: in order to achieve low training loss, the model needs to perform well on the DNS data, as well as, their random orthogonal transformations. To enable sampling, we parametrize the orthogonal transformation as the product of a rotation matrix $\mathbf{R}_{\boldsymbol{\theta}}$ and a reflection matrix $\mathbf{R}_{\textbf{b}}$:
    \begin{gather}
        \label{eq:orth_transform}
        \textbf{R} = \textbf{R}_{\boldsymbol{\theta}}\textbf{R}_{\textbf{b}} \\
        \textbf{R}_{\boldsymbol{\theta}} = \textbf{R}_x(\theta_x)\textbf{R}_y(\theta_y)\textbf{R}_z(\theta_z), \\
        \textbf{R}_x(\theta_x) =
        \begin{bmatrix}
            1 & 0 & 0 \\ 0 & \cos{\theta_x} & \sin{\theta_x} \\0 & -\sin{\theta_x}& \cos{\theta_x} 
        \end{bmatrix},
        \textbf{R}_y(\theta_y) =
        \begin{bmatrix}
            \cos{\theta_y} & 0 & \sin{\theta_y} \\ 0 & 1 & 0 \\ -\sin{\theta_y}& 0 & \cos{\theta_y} 
        \end{bmatrix},
        \textbf{R}_z(\theta_z) =
        \begin{bmatrix}
            \cos{\theta_z} & \sin{\theta_z} & 0 \\ -\sin{\theta_z} & \cos{\theta_z} & 0 \\0 & 0 & 1 
        \end{bmatrix}\\
        \textbf{R}_\text{b} = b\textbf{I} = 
        \begin{bmatrix}
            b & 0 & 0 \\ 0 & b & 0 \\ 0 & 0 & b 
        \end{bmatrix}
        \label{eq:reflect}
    \end{gather}
where $\textbf{R}_x, \textbf{R}_y$ and $\textbf{R}_z$ are elementary rotation matrices about $x, y$ and $z$ axes respectively, with rotation angles specified by $\boldsymbol{\theta} = (\theta_x, \theta_y, \theta_z)$. Each elementary rotation corresponds to rotating a vector in $\mathbb{R}^3$ around the associated axis counter-clockwise or equivalently, the coordinate axes in the opposite direction by the same angle. $\mathbf{R}_{\text{b}}$ is parametrized by a binary variable $b$, which takes the value of $+1$ and $-1$ with equal probability (the latter signifies a reflection). For $\theta_x, \theta_y, \theta_z \in [0, 2\pi]$, (\ref{eq:orth_transform}) to (\ref{eq:reflect}) encompass all possible orthogonal transformations with determinant $+1$ (proper rotation) and $-1$ (improper rotation). Note that such a parametrization is by no means unique. However, it has the fewest number of parameters possible ( 3 angles plus a single binary), offers good geometric interpretability and flexible control over the range of transformation, choosing anywhere between small perturbation and a completely free rotation.

By applying the same transformation, e.g. first order tensors as $\mathcal{R}(\mathbf{v}) = \mathbf{Rv}$ and second order tensors as $\mathcal{R}(\nabla\textbf{u}) = \textbf{R}\nabla\textbf{u} \textbf{R}^{-1}$, to all quantities including
bubble velocity $\bar{\textbf{v}}$ and flow field variables selected to make up $\boldsymbol{\xi}$ along a trajectory, we obtain a rotated trajectory
    \begin{equation}
        \mathcal{T}^{(n)}_{\mathcal{R}} = \{\mathcal{R}(\bar{\textbf{v}}_1^{(n)}), \mathcal{R}(\hat{\boldsymbol{\xi}}_1^{(n)}),...,\mathcal{R}(\bar{\textbf{v}}_T^{(n)}), \mathcal{R}(\hat{\boldsymbol{\xi}}_T^{(n)})\}.
\end{equation}
In this way, for each directly observed trajectory $\mathcal{T}^{(n)}$ we easily obtain an augmented set $\{\mathcal{T}^{(n)}_{1},...,\mathcal{T}^{(n)}_{q}\}$ by applying $q$ different sets of orthogonal transformation parameters $\{(\boldsymbol{\theta}_1, \textbf{b}_1), ..., (\boldsymbol{\theta}_q, \textbf{b}_q)\}$. We refer to $q$ as the \textit{augmentation factor}. We sample all $\boldsymbol{\theta}$ from a normal distribution: $\mathcal{N}(\mathbf{0}, \sigma^2\mathbf{I}), \sigma=0.5$ for each trajectory in the directly observed data set $\mathcal{D}_{\text{o}}$ to obtain an augmented data set, which we denote by $\mathcal{D}_{\text{aug}}$. This sampling regime is found to significantly improve the models while maintaining stable training, as it is demonstrated by the numerical experiments described in the following sections.

\subsection{Cross-validation numerical experiments}

\begin{figure}[ht]
        \centering
        \includegraphics[width=.9\linewidth]{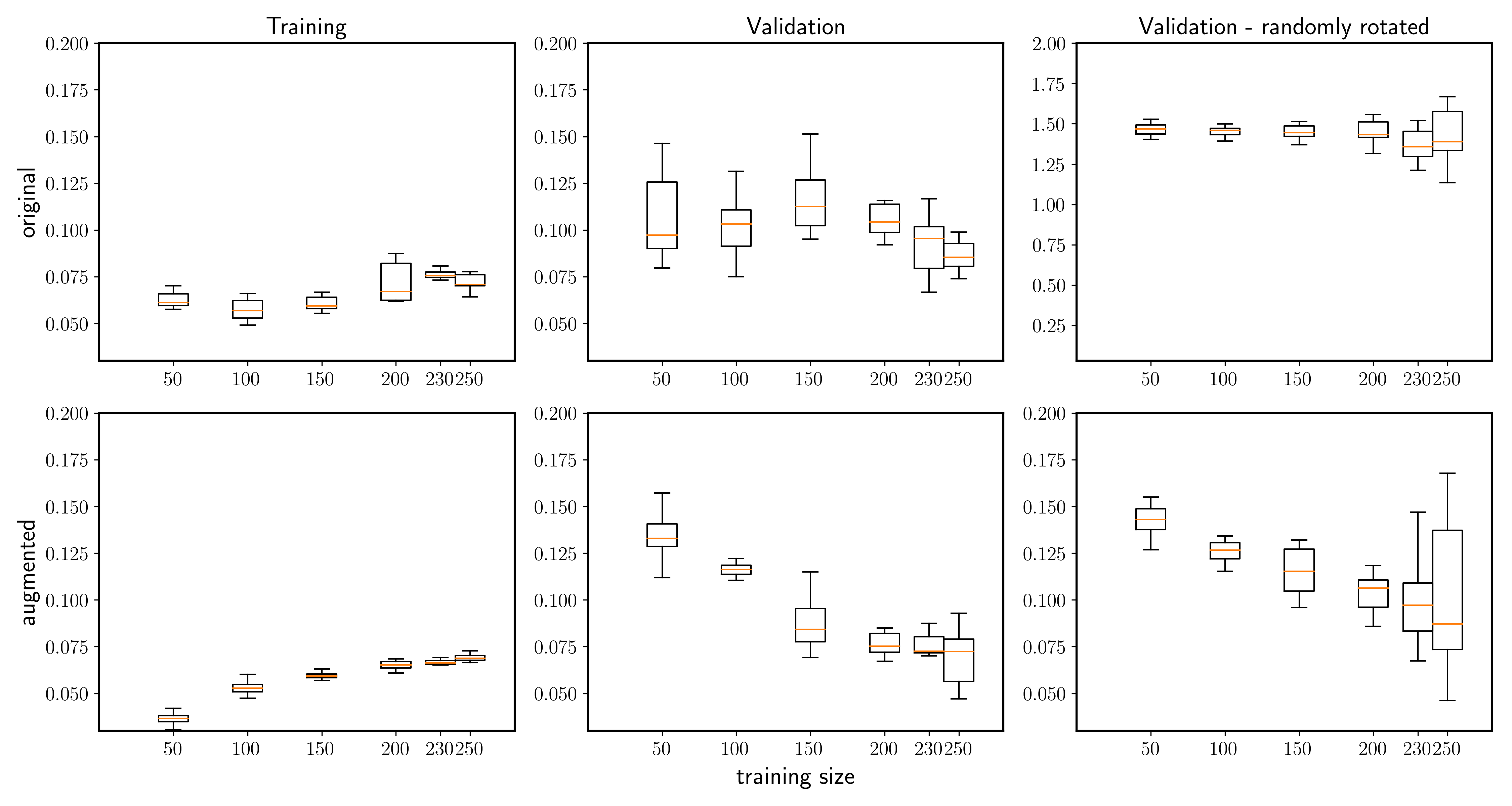}
        \caption{Cross validation box plots showing training and validation MSE against training data size. Each box-whisker represents statistics of 10 independent training-test partitions. Top row model is trained with $\mathcal{D}_{\text{o}}^{\text{train}}$. Bottom row model is trained with $\mathcal{D}_{\text{aug}}^{\text{train}}$. Horizontal axes indicate size of $\mathcal{D}_{\text{o}}^{\text{train}}$ (subtracted from 255 to obtain size of $\mathcal{D}_{\text{o}}^{\text{test}}$); $\mathcal{D}_{\text{aug}}^{\text{train}}$ is augmented from $\mathcal{D}_{\text{o}}^{\text{train}}$ to 8000 trajectory examples (different $q$ for different $\mathcal{D}_{\text{o}}^{\text{train}}$ size). Middle and right columns show MSE computed on $\mathcal{D}_{\text{o}}^{\text{test}}$ and $\mathcal{D}_{\text{aug}}^{\text{test}}$ (augmented from $\mathcal{D}_{\text{o}}^{\text{test}}$ to 1000 trajectory examples) for each trained model respectively. All models use six-dimensional feature vector $\boldsymbol{\xi} = [\textbf{u}, D\textbf{u}/Dt]$. Reynolds number is set at 800.}
        \label{fig:boxplot}
 \end{figure}

We study the effects of training data size as well as the performance of the augmentation scheme through several cross-validation tests. We use as baseline data $\mathcal{D}_{\text{o}}$ the 255 trajectories generated from DNS for $\text{Re}=800$. Each trajectory consists of 400 time steps ($\Delta t = 0.05$). The trajectories are randomly partitioned into a training $\mathcal{D}_{\text{o}}^{\text{train}}$ and a test set $\mathcal{D}_{\text{o}}^{\text{test}}$, at different size ratios. 10 independent partitions are drawn for each training-test ratio. Separate models are then trained and tested on each division of data and the resulting MSE statistics are shown in the form of boxplots in Figure \ref{fig:boxplot}. In the figure we compare performance of model trained from the original $\mathcal{D}_{\text{o}}^{\text{train}}$ and the augmented $\mathcal{D}_{\text{aug}}^{\text{train}}$ on both $\mathcal{D}_{\text{o}}^{\text{test}}$ and $\mathcal{D}_{\text{aug}}^{\text{test}}$. We observe that model trained using $\mathcal{D}_{\text{o}}^{\text{train}}$ does not perform well on $\mathcal{D}_{\text{aug}}^{\text{test}}$, which suggests that the model has developed a preference for axes orientation during training. 

On the other hand, the model trained using $\mathcal{D}_{\text{aug}}^{\text{train}}$ not only generalizes well to both $\mathcal{D}_{\text{o}}^{\text{test}}$ and $\mathcal{D}_{\text{aug}}^{\text{test}}$ but also has better stability (smaller variance) in training error. Note that the larger testing error variance as the size of $\mathcal{D}_{\text{o}}^{\text{train}}$ increases may be attributed to $\mathcal{D}_{\text{o}}^{\text{test}}$ getting extremely small in size and
more likely to sample and overweight extreme examples. Nevertheless, the error mean steadily decreases with richer training data, demonstrating effectiveness of the data augmentation procedure on training directionless models.

\subsection{Model generalization with respect to Reynolds number}
 
It is demonstrated in \cite{wan_sapsis_2018} that a data-driven model can be directly generalized to unseen flow fields, provided that a globally attracting slow manifold exists at all times for the bubble kinematics. This condition is not necessarily satisfied for the DNS training set used in this model. Therefore, we conduct numerical experiments to examine the generalization capacity of the learned model to flow conditions other than those experienced during the training phase. In particular, we focus on applying the model to various Re numbers.

We obtain DNS data for Re = 400, 800 and 1600 respectively. Models (with random sampled and augmented training data) using 6-dimensional $\boldsymbol{\xi} = [\textbf{u}, D\textbf{u}/Dt]$ are learned for each Re number and tested on all others. The MSE results are plotted against time in Figure \ref{fig:MSEcomp}. Time-averaged values of the MSE for each case are given in Appendix B. Each subplot represents comparison of data with the models for a specific Re. Each curve within a subplot represents the error for a particular data-driven model, learned from a given Re data-set or using all Re data-sets together (Re mixed). We also include results from the slow manifold of the MR equation (\ref{eq:MRsm}) (MRsm), as well as, integrating the full MR equation (\ref{eq:MRfull}),
\begin{equation}
        \textbf{v}(t) = \textbf{v}(0) + \int_0^t \frac{1}{\epsilon}(\textbf{u}(\textbf{x}, s) -\textbf{v}(\textbf{x}, s)) + \frac{3R}{2}\frac{D}{Dt}\textbf{u}(\textbf{x}, s)\; ds,
\end{equation} 
where the integral is computed over the \textit{true} DNS trajectory. Both models are far less effective in capturing the bubble velocity compared to the data-driven models.

We note that learning kinematic models for higher Re number is intrinsically more difficult because small scales of the flow become more important and  averaging the input variables $\hat{\xi}$ is less representative of the true flow state in the vicinity of the bubble. This is consistent with the observation that training errors for high Re number flows are typically larger. Nevertheless, when applied to flows with different Re numbers than those used for training, we observe that the models still generate reasonably accurate predictions. As expected, the best performance occurs at the Re number on which they are trained (see also Appendix B). 

\begin{figure}[ht]
        \centering
        \includegraphics[width=0.7\linewidth]{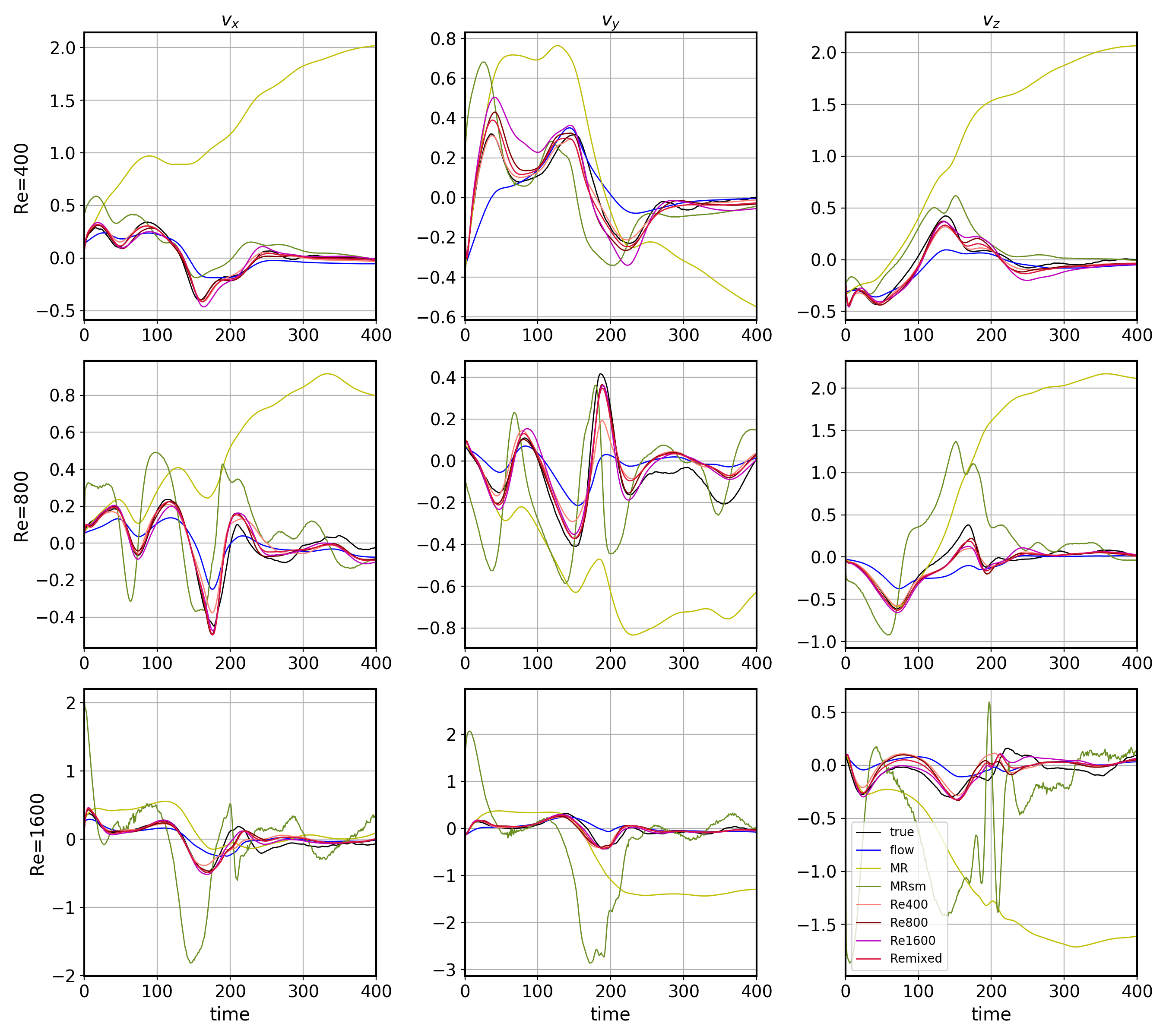}
        \caption{Example predictions for Re = 400, 800 and 1600 trajectories using models trained from data sets with various Re numbers and analytical models, such as MR and the reduced MR. These are superimposed with the  true (DNS) bubble velocity, and the flow velocity. Predictions are bagged results obtained from 20 models (each model trained with different samples from the same data-set).}
        \label{fig:traj_crossRe}
    \end{figure}

We also note that \textit{models trained on higher Re number data sets tend to predict better the trajectories for lower Re flows compared with the model performance on higher Re flows when trained on lower Re data sets}. This is consistent with the fact that higher Re flows are much richer in terms of scales and dynamics. In Figure \ref{fig:traj_crossRe} we plot the predicted velocity time series at each Re number. Specifically, we show velocity predictions for a specific bubble trajectory. These predictions are made with models trained using data sets with various Re numbers. These are compared with the true (DNS) bubble velocity time series, the flow velocity time series, as well as the MR and reduced MR predictions. Data-driven predictions are averaged results obtained from 20 models (each model trained with different samples from the same data-set). We observe that the data-driven models provide much better predictions than MR and notably  for high Re numbers.
 
    \begin{figure}[ht]
        \centering
        \includegraphics[width=0.8\linewidth]{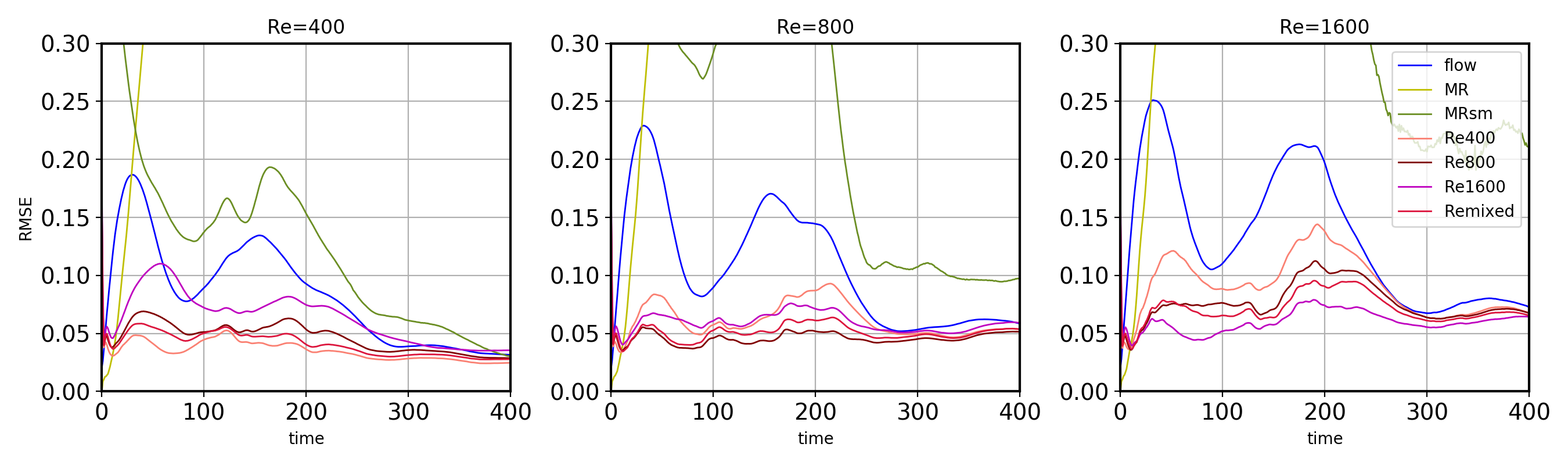}
        \caption{RMSE (unnormalized) vs. time averaged over 5000 augmented test trajectories for Re = 400, 800 and 1600 respectively. The title of each subplot indicates the data-set at which we compare the different models. The different curves indicate different models trained using DNS trajectories for various Re, analytical (such as full MR or slow-manifold of MR), and the fluid flow itself.}
        \label{fig:MSEcomp}
    \end{figure}
 
\section{Application: Steady Hill's spherical vortex}
\label{sec:hills}

We illustrate the performance of the kinematic model, learned from the DNS of the Taylor-Green vortex,  in  the Hill's spherical vortex. Due to the laminar and steady character of this flow, bubbles described by MR tend to cluster in specific regions having the form of limit-cycles. It is therefore interesting to examine whether the data-driven model is able predict the existence of similar limit-cycles that act as attractors for bubbles. 

\begin{figure}[hp]
        \centering
        \begin{subfigure}{.48\textwidth}
                \includegraphics[width=\textwidth]{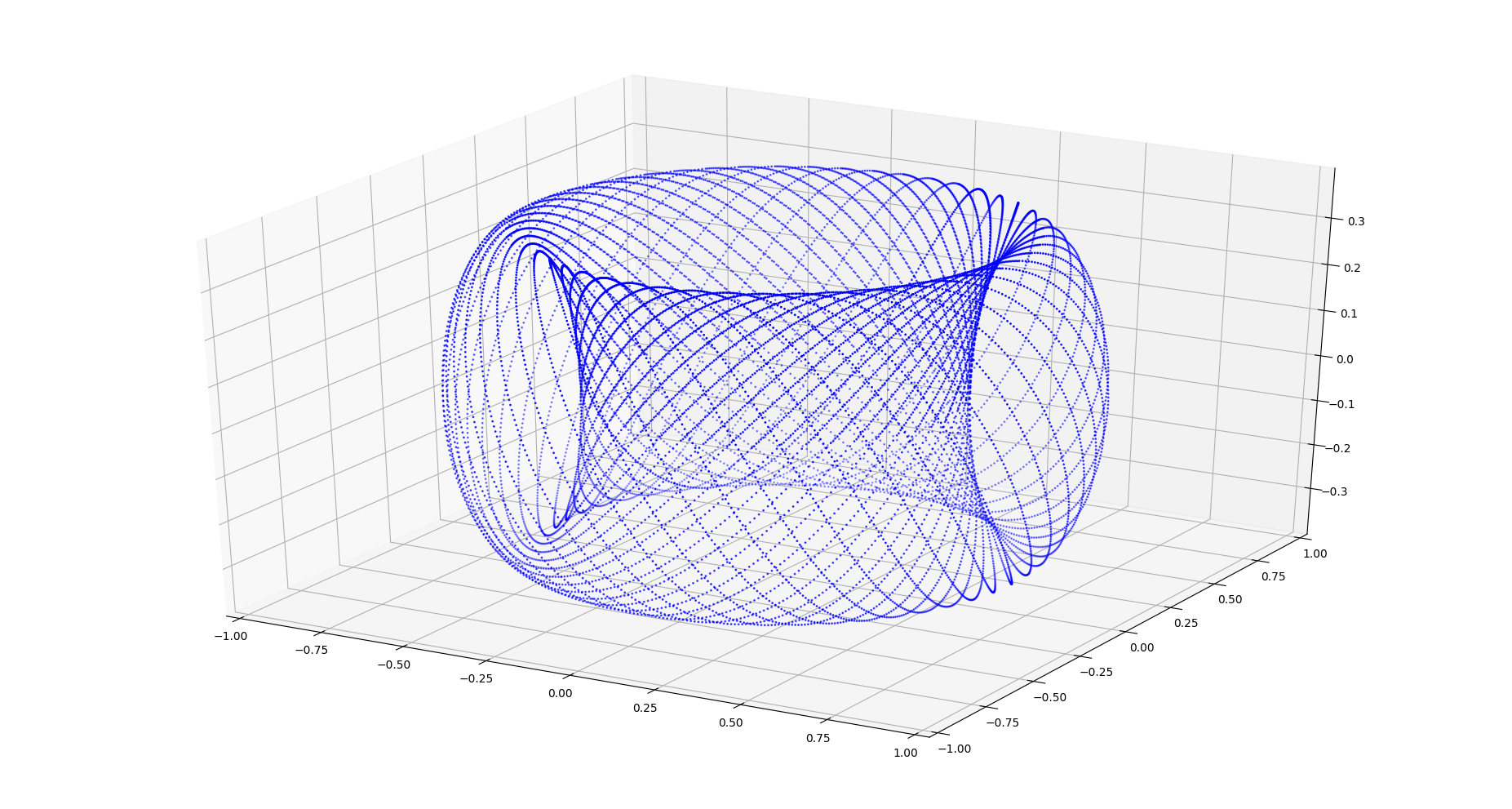}
        \end{subfigure}
        \begin{subfigure}{.45\textwidth}
                \includegraphics[width=\textwidth]{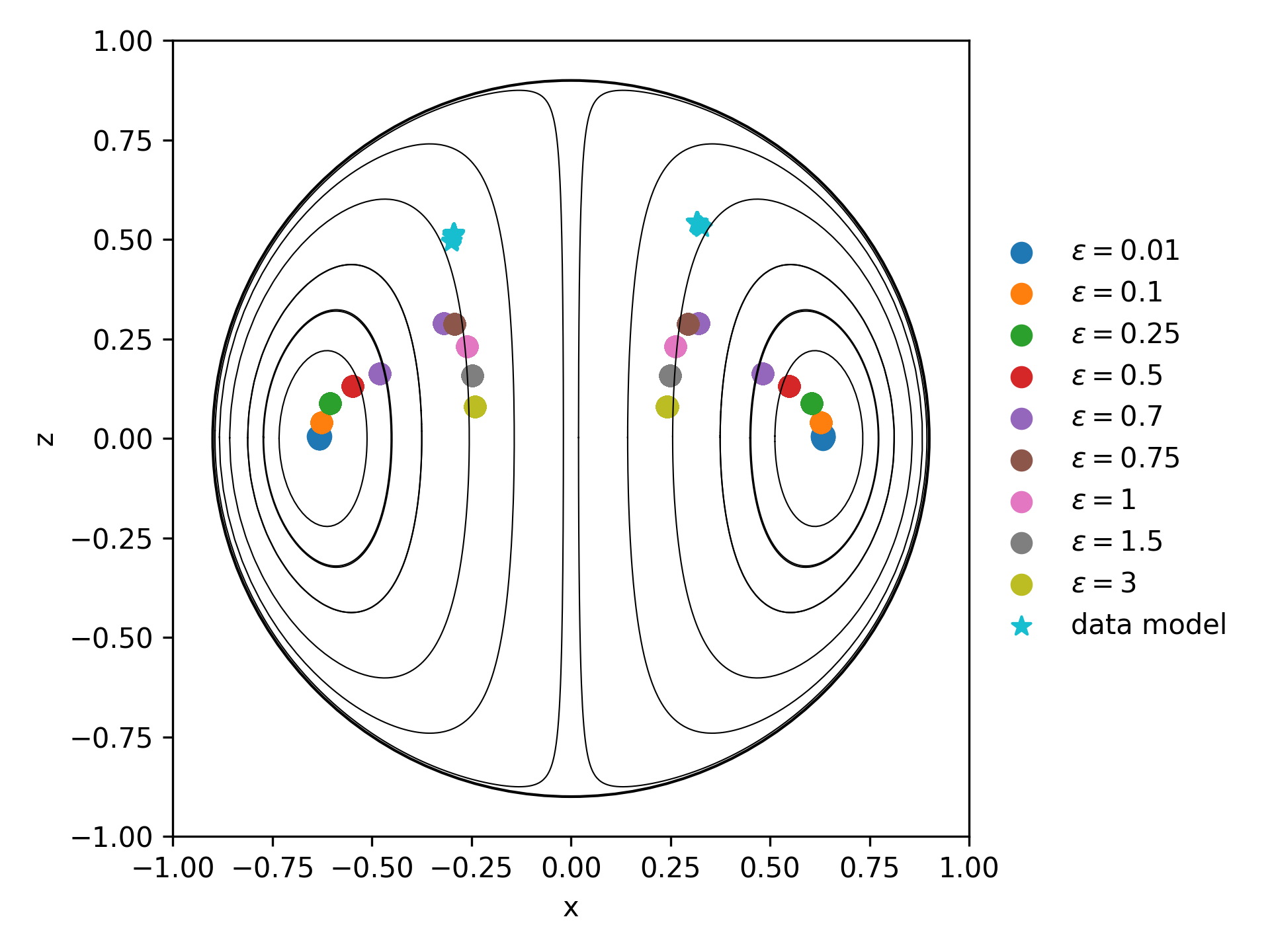}
        \end{subfigure}
        \begin{subfigure}{.45\textwidth}
                \includegraphics[width=\textwidth]{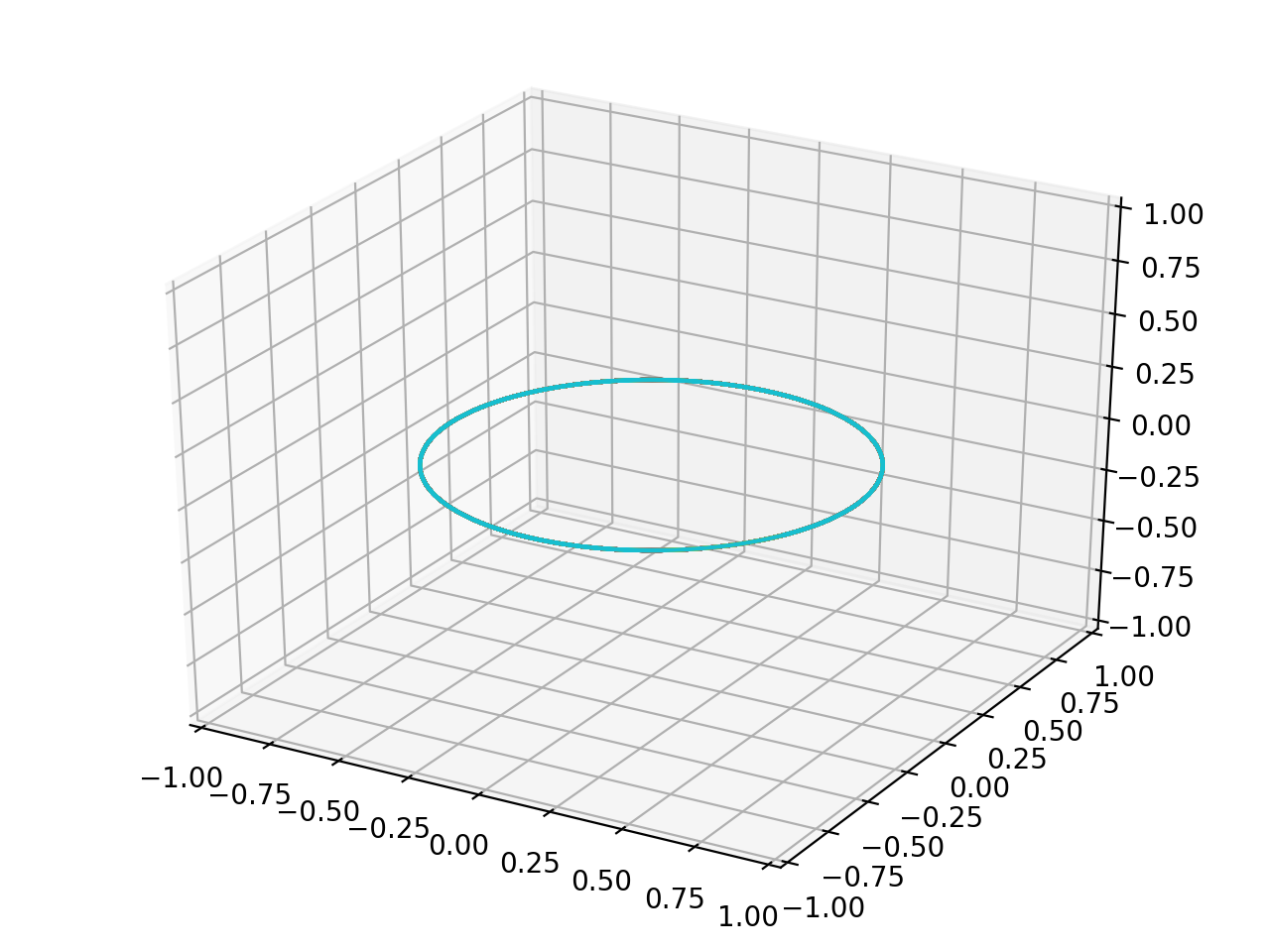}
        \end{subfigure}
        \begin{subfigure}{.45\textwidth}
                \includegraphics[width=\textwidth]{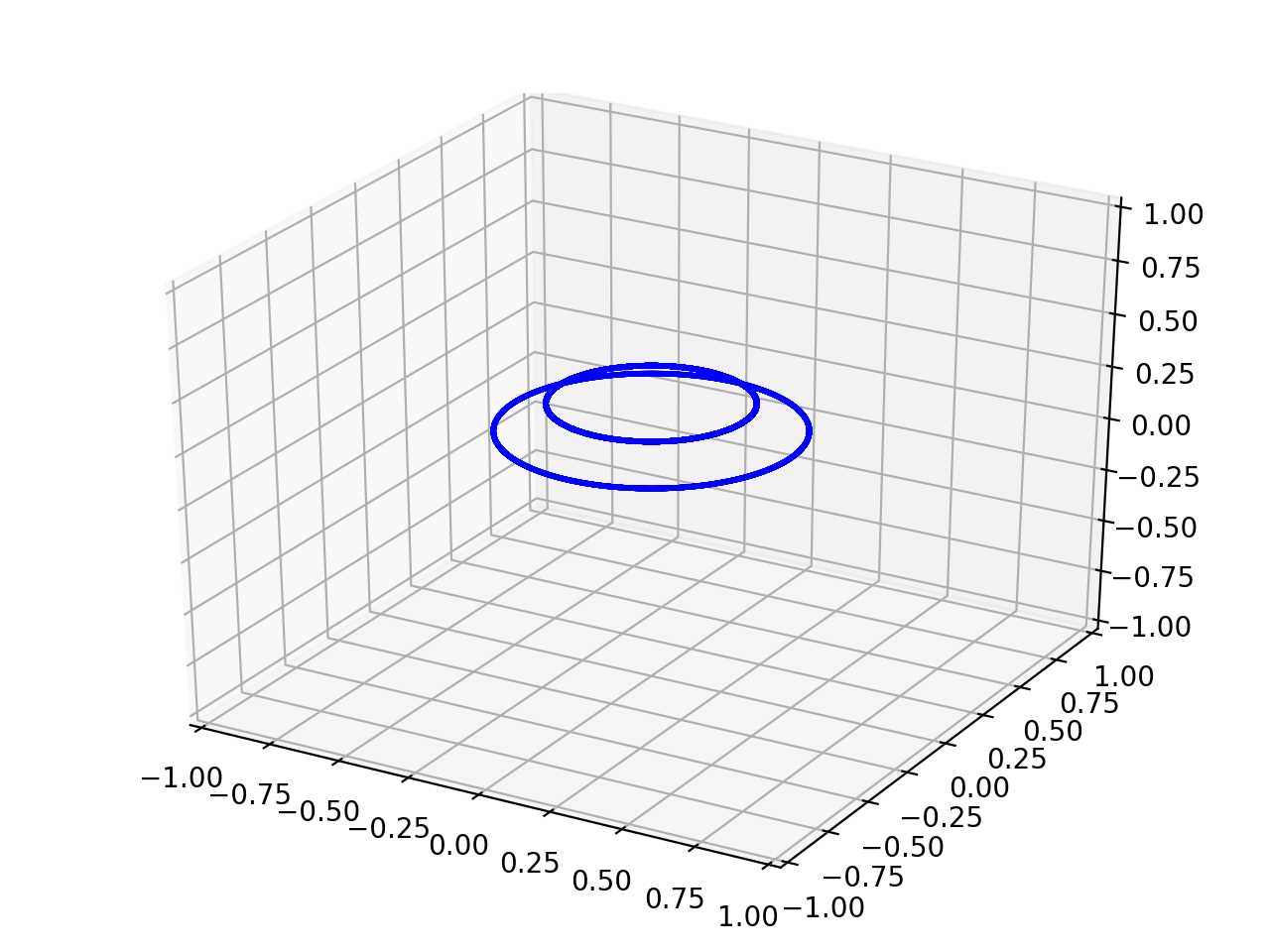}
        \end{subfigure}
        \caption{(a) Toroidal stream surfaces for fluid elements. (b) Poincar\'{e} section at $y=0$ of the Hill's spherical vortex for $a=0.6$, $b=0.1$ and $c=0.15$. Stream surfaces are marked in black. (c) Circular limit cycle generated by the MR equation for $\epsilon=0.01$. (d) A double limit cycle generated by the MR equation for $\epsilon=0.7$.}
        \label{fig:hills_poincare}
 \end{figure}

We consider the integrable case of a Hill's spherical vortex amended with a line vortex at the $z$-axis. The velocity field of the flow is given by
\begin{equation}
        \textbf{u}(\textbf{x}) =
        a\begin{bmatrix}
                xz - 2cy/(x^2+y^2+b) \\
                yz + 2cx/(x^2+y^2+b) \\
                1 - 2(x^2+y^2+b) -z^2
        \end{bmatrix}, 
\end{equation} where $a$ is a scaling parameter 
$b$ is a parameter used to remove the singularity along the $z$-axis. $c$ balances the components (singular and non-singular) in $x$- and $y$-directions. For our study we choose $a = 0.6, b= 0.1$ and $c = 0.15$. This flow generates compact toroidal stream surfaces inside the sphere $|\textbf{x}|\leq1-b$ (see Figure \ref{fig:hills_poincare}a).

We examine the statistical steady state of bubbles following the MR dynamics for a fixed set of flow parameters at different bubbles size ($\epsilon$). To achieve this we randomly pick a number of initial positions inside the spherical vortex. The locations of the bubbles are then evolved based on the MR equation and the machine-learned model. According to the MR dynamics, for sufficiently small $\epsilon$, the bubbles should always cluster on a limit cycle in the shape of a circular ring \cite{Sapsis2010}. However, the position and size of the ring differs for particles with different inertia. For $\epsilon$ close to 0, the MR limit cycle lies on the $z=0$ plane. As $\epsilon$ increases, the MR limit cycles lies higher up along the $z$-axis. At around $\epsilon=0.7$, the system goes through a bifurcation as two stable limit cycles simultaneously exist at different $z$ levels. Of course, this is a regime where the validity of MR is questionable even for simple laminar flows.  As $\epsilon$ increases further, the system returns to having a single stable limit cycle whose vertical position, $z$, and radius nonetheless start to decrease with respect to $\epsilon$. 

\begin{figure}[h]
        \centering
        \begin{subfigure}{.24\textwidth}
                \includegraphics[width=\textwidth]{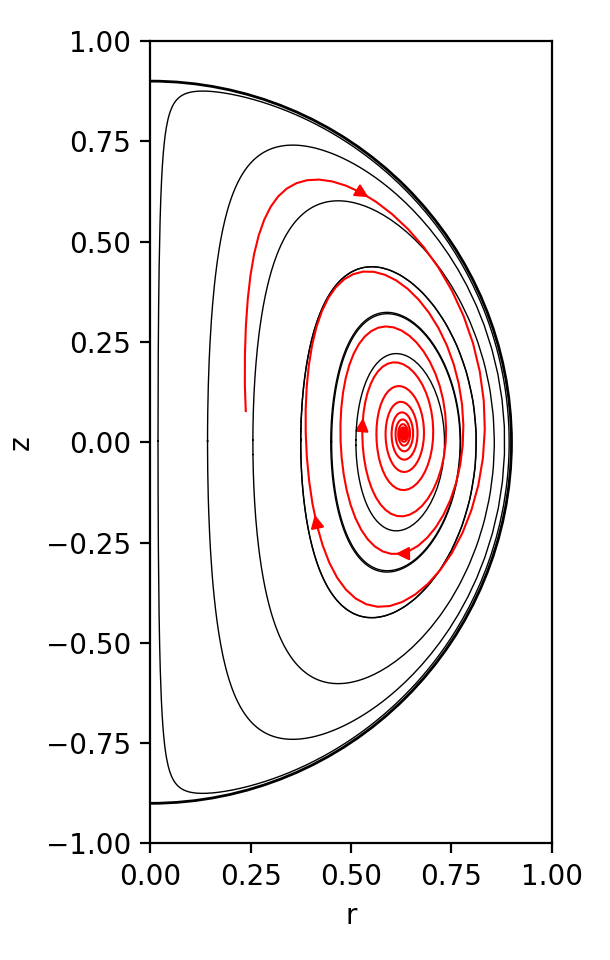}
                \caption{$\epsilon=0.05$}
        \end{subfigure}
        \begin{subfigure}{.24\textwidth}
                \includegraphics[width=\textwidth]{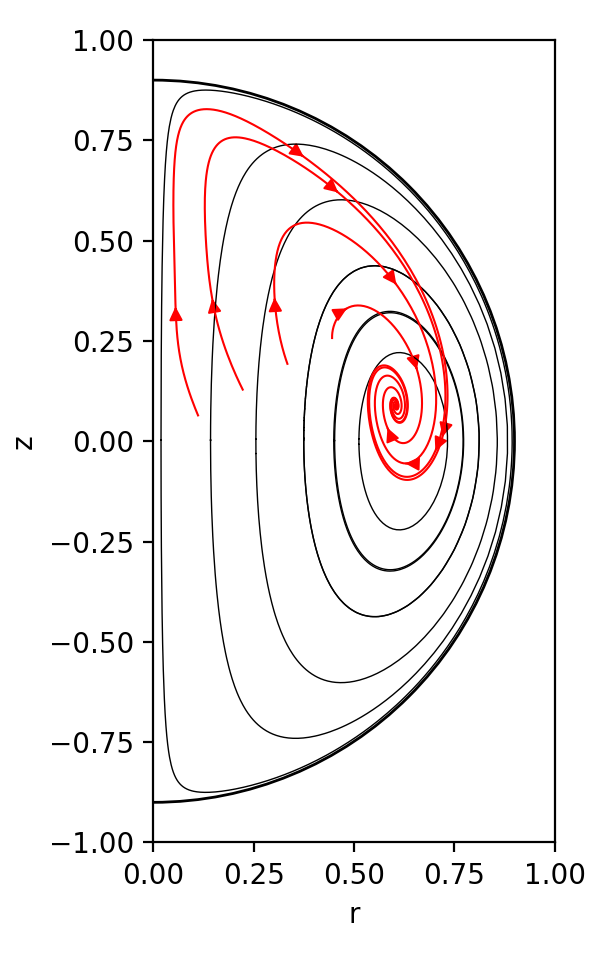}
                \caption{$\epsilon=0.25$}
        \end{subfigure}
        \begin{subfigure}{.24\textwidth}
                \includegraphics[width=\textwidth]{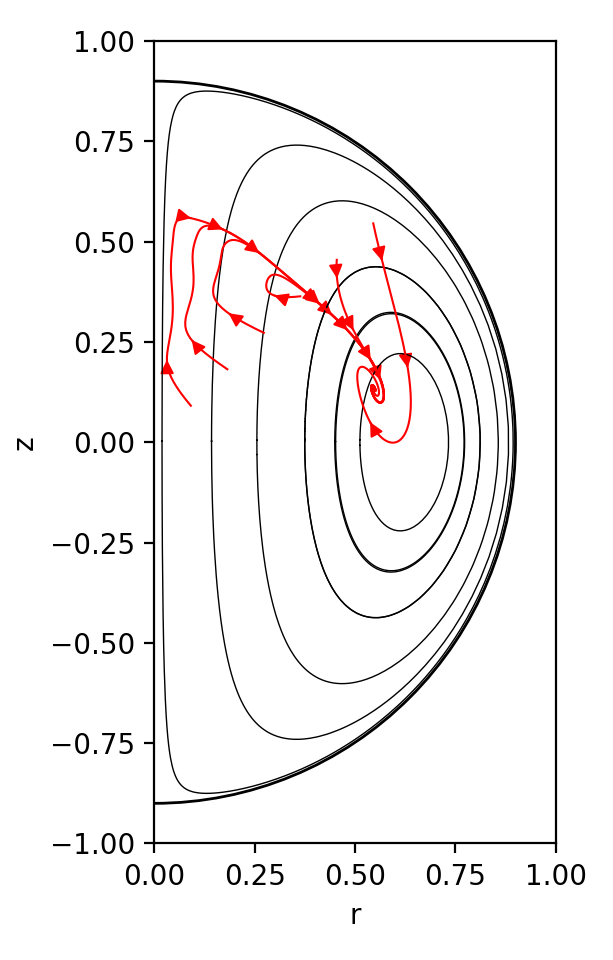}
                \caption{$\epsilon=0.50$}
        \end{subfigure}
        \begin{subfigure}{.24\textwidth}
                \includegraphics[width=\textwidth]{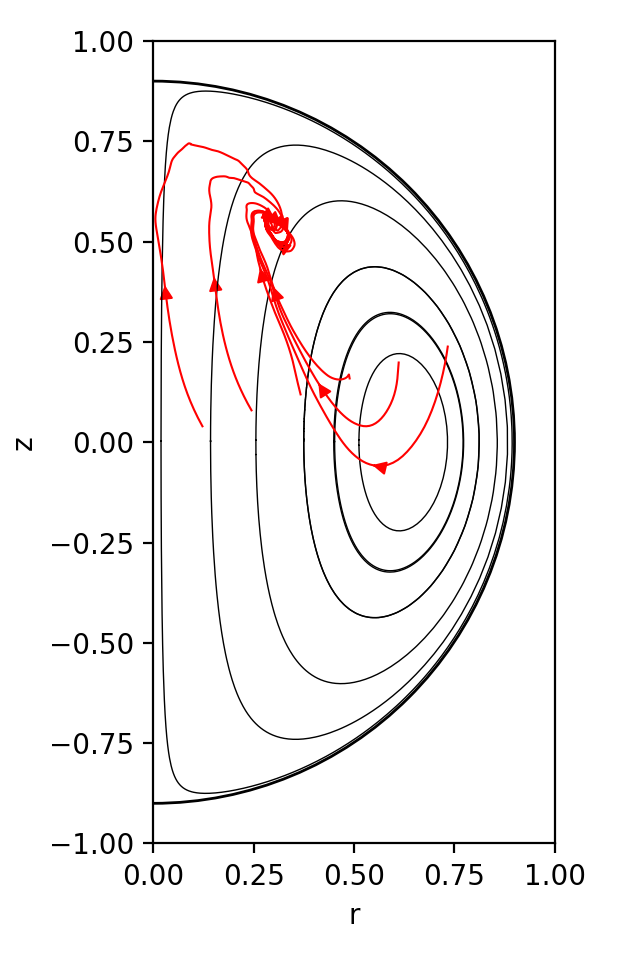}
                \caption{data model}
        \end{subfigure}
        \caption{Trajectory in $(r, z)$ plane (where $r = \sqrt{x^2+y^2}$, corresponding to the plane formed by the $z$-axis and the bubble coordinate) for MR equation using increasing values of $\epsilon$ (a to c) and for the data model (d) which corresponds to $\epsilon=1.5$ approximately.}
        \label{fig:hills_fixpts}
 \end{figure}

The data model on the other hand describes the motion of a bubble with large size. For the purpose of comparison, we set original scales of the Hill's vortex flow to be the same as those of the DNS, which leads to $\epsilon=1.5$. In this case the limit cycle position is even higher up along the $z$-axis compared to all MR cases. The overall trend is illustrated by the Poincar\'{e} map in Figure \ref{fig:hills_poincare}b and also the plots in Figure \ref{fig:hills_fixpts}. It is interesting to note that the two approaches agree qualitatively on the existence of a limit cycle. Quantitatively, however, the data-driven method predicts a different location for the limit cycle, which is consistent with the trend that the MR gives for small $\epsilon$, i.e. for values $0-0.5,$ where we expect it to be valid.
 
\section{Application: Unsteady vortex ring}
\label{sec:ring}

\begin{figure}[b]
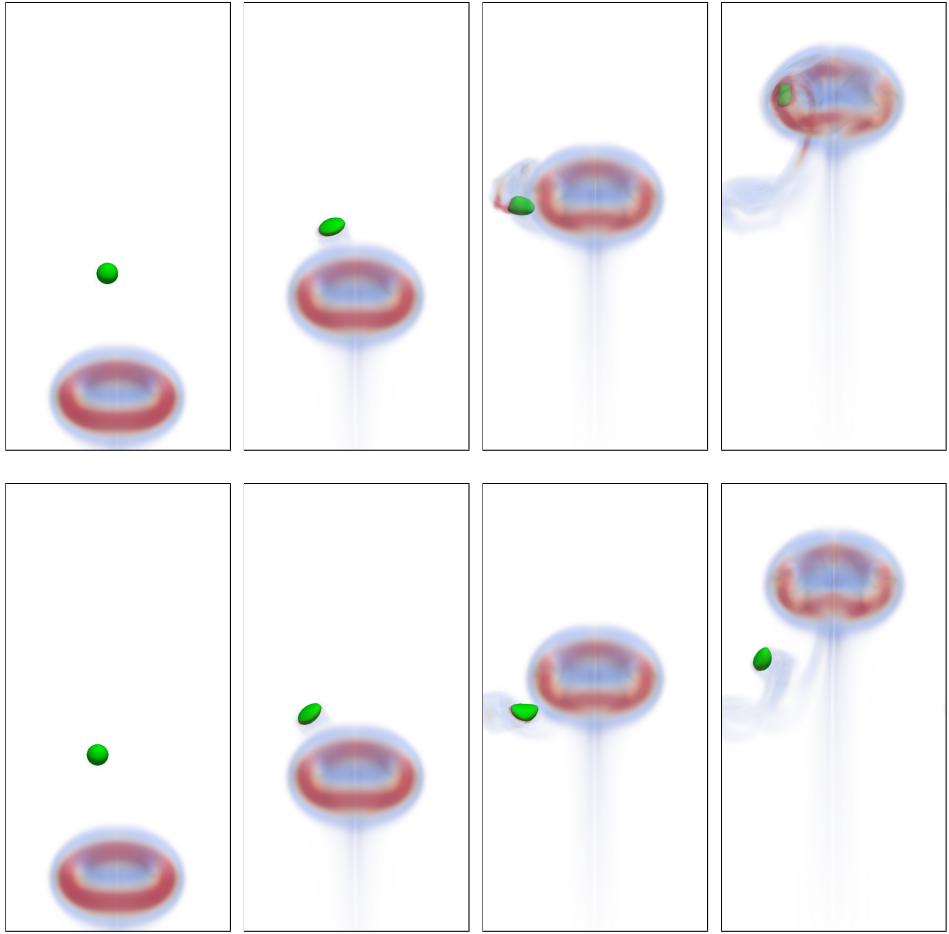

    \centering
    \newcommand{\p}[2]{%
      \hspace{1mm}%
      \frame{\includegraphics[width=0.18\textwidth]{b#1_c0#2}}%
      \hspace{1mm}%
    }

    \p{07}{040}%
    \p{07}{060}%
    \p{07}{080}%
    \p{07}{100}

    \bigskip

    \p{05}{040}%
    \p{05}{060}%
    \p{05}{080}%
    \p{05}{100}

    \caption{
      Interaction of a vortex ring with a bubble.
      Snapshots of the vorticity magnitude (increasing values from blue to red)
      and the bubble surface (green)
      at $t=40,\;60,\;80$ and $100$.
      Parameters of the vortex ring are given in \cite{bergdorf2007} (ring A)
      and the ring radius is $R=1$.
      Bubble of radius $R_b=0.25$ is positioned at a distance $\Delta$
      from the ring axis.
      Depending on~$\Delta$, 
      the bubble is either trapped inside the ring (top, $\Delta=0.22$)
      or separates from the ring (bottom, $\Delta=0.44$).
    }
    \label{fig:ring}
 \end{figure}

As a last example, we examine the performance of the data driven, RNN model in an unsteady flow, generated by DNS of the NS equations. The DNS  resolve a bubble whose motion is driven by an evolving vortex ring. The  vortex ring is initialized  by a Gaussian azimuthal vorticity in cylindrical coordinates
\begin{equation}
    \omega_\theta(\mathbf{x}, t=0) = \frac{1}{\pi\sigma^2}e^{-(s/\sigma)^2},
\end{equation}
where $s^2=z^2+(\sqrt{x^2+y^2} - 1)^2$ is the squared distance away from the circular ring $\{\Omega_R: x^2+y^2=1, z=0\}$ that is the location of maximum vorticity and also the mean of the Gaussian. The standard deviation is denoted with $\sigma$ and its value is chosen as $\sigma=2.4232$ \cite{bergdorf2007}. We use the Stokes stream function to solve for the initial velocity field and evolve the flow according to (\ref{eq:NS}) and (\ref{eq:volfrac}). The results are shown in Figure \ref{fig:ring}. For $t > 0$, the vortex ring  moves in the positive $z$-direction, steadily, having a constant velocity. A bubble that is initially resting above the vortex ring interacts with the flow in two possible ways as the ring passes by: (a) the bubble becomes trapped inside the ring and moves together with it or (b) the bubble slips away. Given the fixed physical properties of the bubble, occurrence of trapping within the vortex ring depends on how far the bubble initial position is chosen from the $z$-axis.

\begin{figure}[t]
        \centering
        \includegraphics[width=\textwidth]{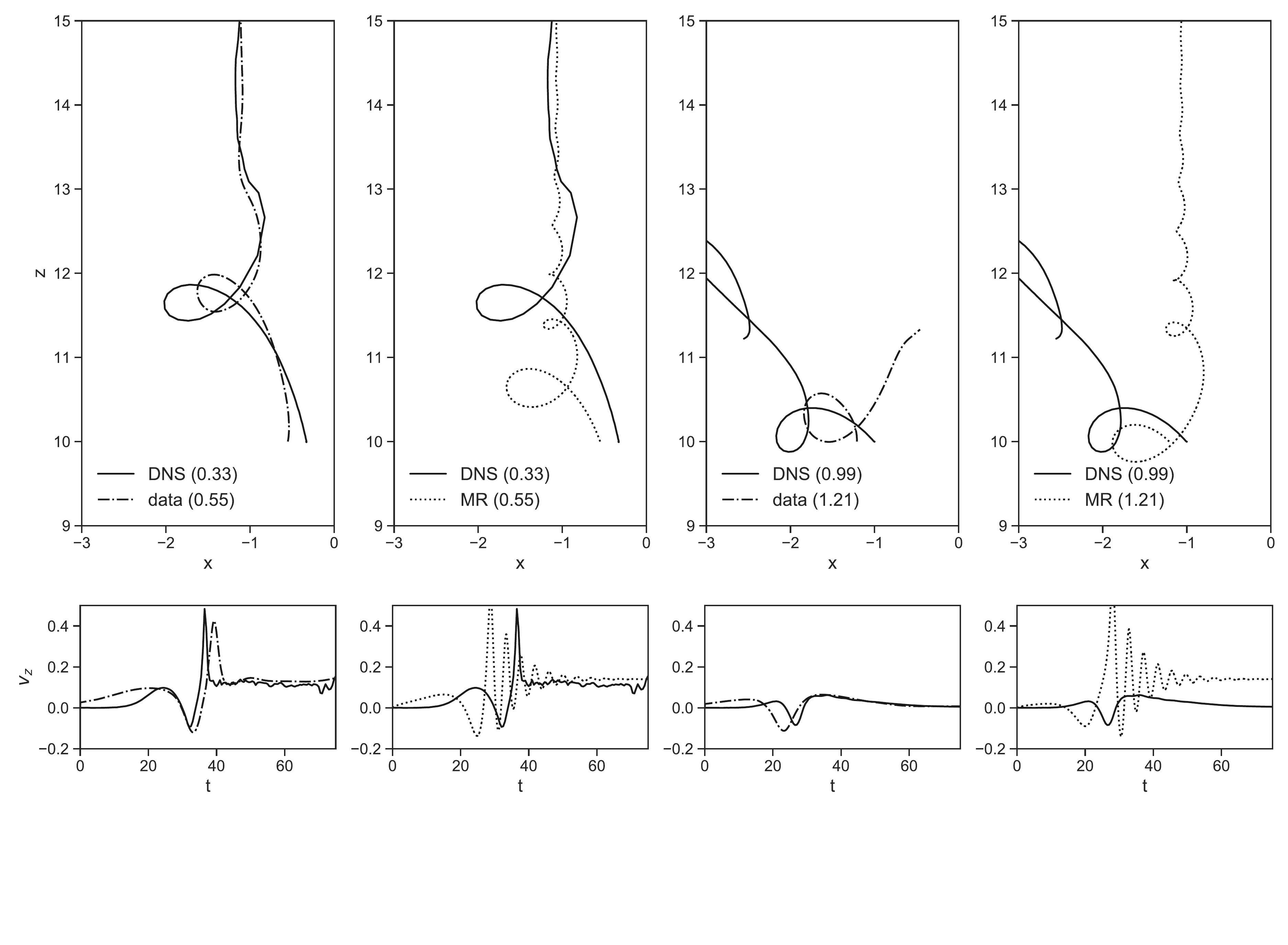}\hfill
        \caption{Comparison of trajectories for DNS, data-driven and MR models. Left two columns correspond to a bubble trapping case within the ring (bubble moves together with traveling vortex ring up and beyond $z=15$) and right two columns are for a slipping case (bubble does not move up after contact with the vortex). Bottom row are the corresponding velocity time series for the first 75 time units. Numbers inside brackets of the legend entries indicate the initial distance away from the $z$-axis. 
        }
        \label{fig:ringtraj}
 \end{figure}

Using the previously learned kinematic model, we track this bubble-flow interaction, solving a simple ordinary differential equation. Specifically, we employ the kinematic model trained for $\text{Re}=1600$, as it is the closest to $Re=7500$ that is used to set up the flow field and DNS for the considered flow. By approximating the flow as steady, in the frame of reference of the vortex ring, input for the data-driven model may be conveniently queried to facilitate fast multiple-step predictions of bubble trajectories. Figure \ref{fig:ringtraj} compares the DNS and data-driven trajectories for both trapping and slipping scenarios. For the trapping case, the DNS and the data-driven model predict comparable trajectories, as well as, velocity time series for bubbles initiated from similar heights and distances away from the central axis. On the other hand, although the MR model correctly predicts trapping, the predicted trajectory contains much more spurious oscillations. It also fails to predict slipping to happen for set-ups which the data driven model is otherwise able to truthfully reflect. Note that the small difference in the initial positions between DNS and the MR or the data-driven model is because we tried to also match the initial velocity of the bubble as this is important for the MR equation. This difference is much smaller than the DNS bubble size. Moreover, we have examined the robustness of the results shown for small perturbations of the initial bubble positions (i.e. within the DNS bubble region) used in the MR and the date-driven model.


\section{Conclusions}

We  propose a machine learning algorithm for  deriving kinematic models of finite-size bubbles with constant radius, in turbulent flows. The kinematic models rely on Recurrent Neural Networks with Long Short Memory (RNN-LSTM) trained on  trajectories obtained from multi-phase DNS of bubbles in turbulent flows. Our approach relies on representing the vector field that governs the velocity of bubbles, in terms of the flow field encountered all along their trajectories. We have assessed various flow properties as potential input variables for the kinematic model and we found as  most informative the velocity field and the local gradient tensor. This is consistent with the general form of the MR equation and its extensions based on the Basset-Boussinesq memory terms. 

We have also discussed a new data augmentation approach that enforces rotational symmetries that the kinematic model should possess. In this way, we are able to augment the  data obtained from the  from DNS. We  studied the predictive generalization  of the models to different Re numbers and observed that the models trained with flows of a given Re number are able to maintain good performance for flows with lower Re numbers, but less so when the relation is reversed. This is not a surprise given that higher Re flows contain smaller scale turbulent structures which are absent in lower Re flows. We  applied the trained model on a steady laminar Hill's spherical vortex. This is a setup where the MR equation has been studied analytically and for which it has been shown that finite-size effects will induce clustering of bubbles on a limit cycle. The results obtained from the data-driven model also predict the existence of a limit cycle which has geometrical characteristics that are consistent with the ones obtained from MR when the bubble size is assumed small. Finally, we use the trained model to perform multi-step prediction for an unsteady flow - a traveling Gaussian vortex ring. With no additional training, predicted trajectories capture very well the trapping and slipping behaviors observed in the multiphase DNS. 

Our analysis relied on the assumption of bubbles having approximately fixed size. Several important directions will be pursued in the future including modeling the effect of variable bubble size, possibly by combining data-driven kinematic models with the Rayleigh-Plasset equation. However, even under this assumption we expect that the presented data-driven strategy for learning the dynamics of bubbles will  impact  a series of applications where analytical models, such as MR equation, encounter limitations, such as quantifying the statistics of bubbles in multiphase flows, as well as, their effect back to the fluid flow.

\subsubsection*{Acknowledgments}
TPS and ZYW have been supported through the ONR-MURI grant N00014-17-1-2676. PK and PK acknowledge the support from \verb`CRSII5_173860` of the Swiss National Science Foundation and the use of computing resources from CSCS (projects s754 and s931).

\balance
\bibliographystyle{abbrv}
\bibliography{ref,library}

\clearpage
\appendix
\section{Input variables selection}

    \begin{table}[h]
        \centering
        \begin{tabular}{cccc}
            \hline
            Input variables             & Dimension            & Training MSE      & Validation MSE      \\ \hline
            $\textbf{u}$         & 3                    & 0.704             & 0.708                \\
            $\frac{D\textbf{u}}{Dt}$         & 3        & 0.094             & 0.086                \\
            $(\textbf{u} \cdot \nabla) \textbf{u}$         & 3              & 0.725                & 0.801                \\
            $\frac{\partial\textbf{u}}{\partial t}$         & 3             & 0.286                & 0.307                \\
            $\textbf{u}, \frac{D\textbf{u}}{Dt}$         & 6                & 0.078                & 0.082                \\
            $\textbf{u}, \frac{D\textbf{u}}{Dt}, \nabla p$         & 9      & 0.102                & 0.115                \\
            $\textbf{u}, \nabla \textbf{u}, \frac{\partial \textbf{u}}{\partial t}$         & 15                   & 0.182                & 0.289       \\ \hline
\end{tabular}
\caption{Performance for different input variables: training and validation loss at the end of 100 training epochs. All input variables are averaged over a spherical surface that contains the bubble (Figure \ref{fig:flow}). The same two-layer LSTM architecture is used for all runs. All results are for Re = 800.}
\label{tab:feat}
\end{table}

\section{Mean squared error for Cross-Re validation}

\begin{table}[h]
        \centering
        \begin{tabular}{c|cc|cc|cc|cccc}
        \hline
        \multirow{3}{*}{\begin{tabular}[c]{@{}c@{}}Training\\ Data\end{tabular}} & \multicolumn{2}{c|}{\multirow{2}{*}{\begin{tabular}[c]{@{}c@{}}Training\\
        MSE\end{tabular}}} & \multicolumn{4}{c|}{Test MSE, Re = 400} & \multicolumn{4}{c}{Test MSE, Re = 800}  \\ \cline{4-11} & \multicolumn{2}{c|}{} & \multicolumn{2}{c|}{original} & \multicolumn{2}{c|}{augmented} & \multicolumn{2}{c|}{original}        & \multicolumn{2}{c}{augmented} \\
        \cline{2-11} 
& mean                                          & std                   
                      & mean          & std           & mean           &
std           & mean   & \multicolumn{1}{c|}{std}    & mean          & std
          \\ \hline
Re = 400                                                                
& 0.0395                                        & 0.0011                
                      & 0.0577        & 0.0038        & 0.0471         &
0.0023        & 0.1093 & \multicolumn{1}{c|}{0.0027} & 0.0953        & 0.0025
       \\
Re = 800                                                                
& 0.0650                                        & 0.0025                
                      & 0.1523        & 0.0038        & 0.1386         &
0.0031        & 0.0777 & \multicolumn{1}{c|}{0.0036} & 0.0767        & 0.0026
       \\
Re = 1600                                                               
& 0.0866                                        & 0.0036                
                      & 0.2850        & 0.0058        & 0.2453         &
0.0046        & 0.1919 & \multicolumn{1}{c|}{0.0021} & 0.1674        & 0.0027
       \\
Re mixed                                                                
& 0.0887                                        & 0.0018                
                      & 0.1683        & 0.0036        & 0.1451         &
0.0026        & 0.1295 & \multicolumn{1}{c|}{0.0021} & 0.1155        & 0.0018
       \\ \hline
\end{tabular}

\begin{tabular}{c|cc|cc|cc|cccc}
\hline
\multirow{3}{*}{\begin{tabular}[c]{@{}c@{}}Training\\ Data\end{tabular}}
& \multicolumn{2}{c|}{\multirow{2}{*}{\begin{tabular}[c]{@{}c@{}}Training\\
MSE\end{tabular}}} & \multicolumn{4}{c|}{Test MSE, Re = 1600}           
           & \multicolumn{4}{c}{Test MSE, combined}                     
         \\ \cline{4-11} 
                                                                        
& \multicolumn{2}{c|}{}                                                 
                      & \multicolumn{2}{c|}{original} & \multicolumn{2}{c|}{augmented}
& \multicolumn{2}{c|}{original}        & \multicolumn{2}{c}{augmented} \\
\cline{2-11} 
                                                                        
& mean                                          & std                   
                      & mean          & std           & mean           &
std           & mean   & \multicolumn{1}{c|}{std}    & mean          & std
          \\ \hline
Re = 400                                                                
& 0.0395                                        & 0.0011                
                      & 0.1696        & 0.0064        & 0.1523         &
0.0056        & 0.0799 & \multicolumn{1}{c|}{0.0029} & 0.0726        & 0.0015
       \\
Re = 800                                                                
& 0.0650                                        & 0.0025                
                      & 0.1338        & 0.0022        & 0.1253         &
0.0022        & 0.0971 & \multicolumn{1}{c|}{0.0026} & 0.0927        & 0.0016
       \\
Re = 1600                                                               
& 0.0866                                        & 0.0036                
                      & 0.0946        & 0.0032        & 0.0957         &
0.0030        & 0.1535 & \multicolumn{1}{c|}{0.0049} & 0.1417        & 0.0025
       \\
Re mixed                                                                
& 0.0887                                        & 0.0018                
                      & 0.1367        & 0.0029        & 0.1289         &
0.0025        & 0.1094 & \multicolumn{1}{c|}{0.0023} & 0.1015        & 0.0012
       \\ \hline
\end{tabular}

\caption{Cross-Re normalized  MSE results with respect to the bubble velocity st. deviation: models trained using data with a particular Re (first column) are tested on data with other Re numbers. Tests are conducted on both original and augmented data set. For each training Re number, 20 sets of 200 trajectories (except for the mixed Re case, where 20 sets of 600 are used) are independently sampled to form augmented data sets. Standard deviations listed here are calculated for the performance of the corresponding models obtained using these random training samples.}
\label{tab:MSE}
\end{table}

\end{document}